# Engineering One-Dimensional Quantum Stripes from Superlattices of Two-Dimensional Layered Materials

*John H. Gruenewald, Jungho Kim, Heung Sik Kim, Jared M. Johnson, Jinwoo Hwang, Maryam Souri, Jasminka Terzic, Seo Hyoung Chang, Ayman Said, Joseph W. Brill, Gang Cao, Hae-Young Kee, and Sung S. Ambrose Seo\**

J. H. Gruenewald, M. Souri, J. Terzic, Prof. J. W. Brill, Prof. G. Cao, Prof. S. S. A. Seo
Department of Physics and Astronomy, University of Kentucky, Lexington, KY 40506, USA
E-mail: a.seo@uky.edu
Dr. J. Kim, Dr. A. Said
Advanced Photon Source, Argonne National Laboratory, Argonne, IL 60439, USA
Dr. H. S. Kim, Prof. H. Y. Kee
Department of Physics, University of Toronto, Toronto, Ontario M5S 1A7, Canada
J. M. Johnson, Prof. J. Hwang
Department of Materials Science and Engineering, The Ohio State University, Columbus, OH 43210, USA
Dr. S. H. Chang
Materials Science Division, Argonne National Laboratory, Argonne, IL 60439, USA

**Abstract**

One-dimensional (1D) quantum systems, which are predicted to exhibit novel states of matter in theory, have been elusive in experiment. Here we report a superlattice method of creating artificial 1D quantum stripes, which offers dimensional tunability from two- to one-dimensions. As a model system, we have fabricated 1D iridium (Ir) stripes using *a-axis oriented superlattices* of a relativistic Mott insulator $Sr_2IrO_4$ and a wide bandgap insulator $LaSrGaO_4$, both of which are crystals with layered structure. In addition to the successful formation of 1D Ir-stripe structure, we have observed 1D quantum-confined electronic states from optical spectroscopy and resonant inelastic x-ray scattering. Since this 1D superlattice approach can be applied to a wide range of layered materials, it opens a new era of 1D science.

An ideal one-dimensional quantum-stripe consists of an infinite linear chain of bonded atoms, wherein the interactions between each atom's electrons are restricted to the single dimension of the chain.[1] One-dimensional (1D) order spontaneously emerges at the onset of quantum phase transitions in the form of charge/spin density waves in multiferroics, superconductors, and as edge and surface states in topologically non-trivial systems.[2-8] An experimental approach for investigating such phenomena is to dimensionally tune these systems to one dimension until a critical phase transition is reached, as conducted similarly in Ruddlesden-Popper series compounds from three to two dimensions.[9] While many successful strategies have been demonstrated for chemically synthesizing 1D nanostructures, direct atomic-layer control between two and one dimension(s) is generally considered unavailable.[10] Additionally, the quantum confinement of conventionally two dimensional materials to one dimension can reveal hidden electronic and magnetic properties. For example, $SrCuO_2$, i.e. 1D chain compound, shows novel spin-charge separation, which is absent in its 2D counterpart, $Sr_2CuO_2Cl_2$ (Ref. [11]). Despite the promising outlook for 1D materials, experimental progress remains in its infancy, hampered by its reliance on the few materials with intrinsic 1D structures.[12]

Here we present a new approach of synthesizing 1D quantum systems by constructing dimensionally-confined stripe-superlattices from *in-plane* oriented 2D layered crystals. Layered transition metal oxides of the form $A_2BO_4$ consist of 2D layers ($BO_2$) in the *ab*-plane that are stacked along the *c*-axis. Such materials are considered 2D since each consecutive plane is well separated by an electronically inert rock salt layer ($AO$) and offset by a half unit cell. Compounds with this so-termed $K_2NiF_4$-type structure can be epitaxially grown such that the 2D layers are oriented parallel to the surface normal direction, i.e. *a*-axis orientation.[13-15] The crux of our idea is that the dimensionality of the 2D layered materials in this *a*-axis orientation can be tuned by restricting the number of monolayers *m* grown along the *in*-plane direction (Figure 1a). For instance, if only a single monolayer is grown, i.e. *m* = 1, then the 2D planes become 1D stripes. To achieve a volume suitable for experimental characterizations, superlattice structures can be grown consisting of alternating layers of *m* monolayers of $BO_2$ and *n* monolayers of an inert, wide bandgap material also containing the $K_2NiF_4$ symmetry ($A'_2B'O_4$). In this manner, the 2D planes can be incrementally tuned to 1D quantum stripes.

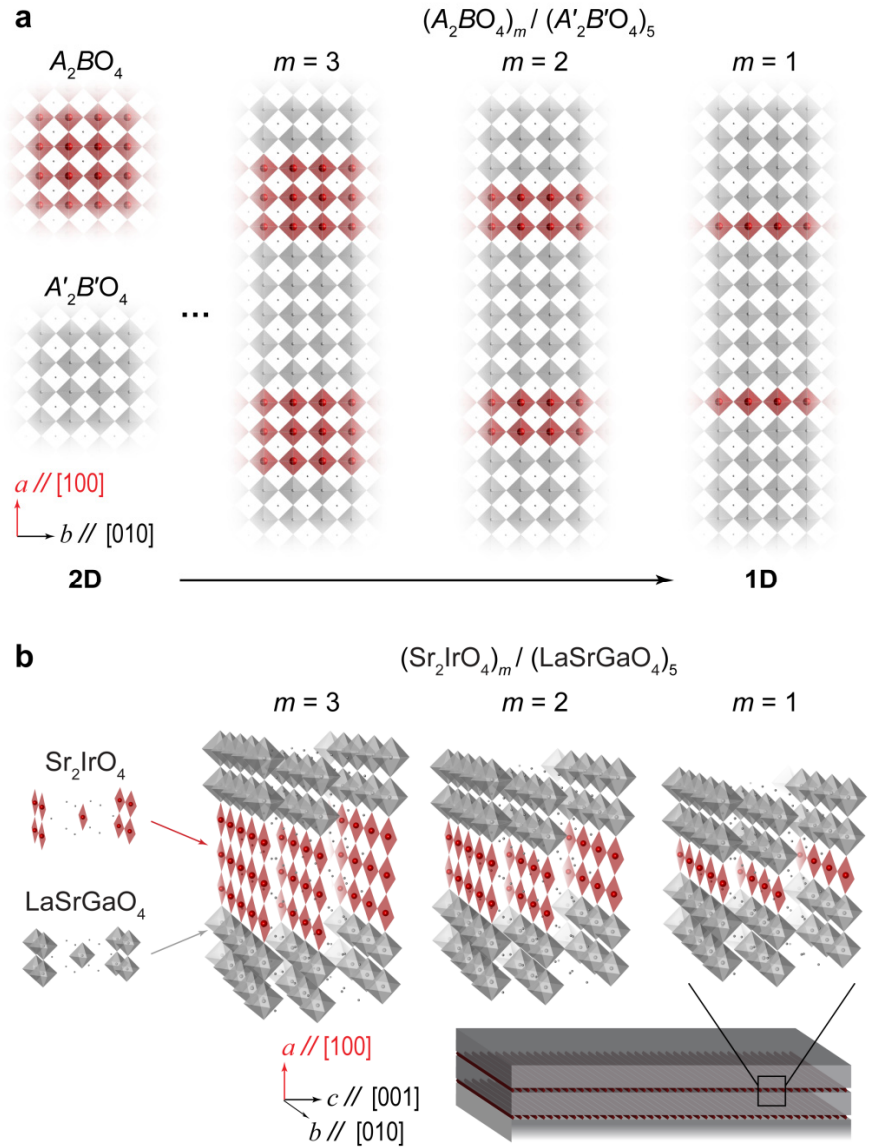

**Figure 1.** Conceptual diagram of turning a 2D layered material into a 1D quantum stripe superlattice. a) The leftmost panel shows the in-plane structures of two transition-metal oxides, $A_2BO_4$ (above) and $A'_2B'O_4$ (below), with the $K_2NiF_4$ symmetry. Each red (gray) square contains transition-metal ions $B$ ($B'$) at its center and an oxygen atom at each of its 4 vertices. The number of $BO_2$ monolayers $m$ corresponds to the horizontal rows of red squares, i.e. the stripes. Low-dimensional stripe phases approaching 1D (e.g. $m$ = 3, 2, 1) can be created by alternating the layers between $m$ monolayers of $BO_2$ and a constant number of $B'O_2$ monolayers (such as 5 in the schematic). The 1D quantum-stripes are achieved in the $m$ = 1 case depicted in the rightmost panel. b) Schematic diagrams of $a$-axis oriented $(Sr_2IrO_4)_m/(LaSrGaO_4)_5$ superlattices for $m$ = 3, $m$ = 2 and $m$ = 1 for realizing the low-dimensional quantum stripes of $IrO_2$ (red squares) on $LaSrGaO_4$ (100) substrates. The 1D $IrO_2$ stripes run parallel to the $b$-axis and are dimensionally confined by the wide bandgap $LaSrGaO_4$ layers (grey octahedra).

We have implemented this idea in synthesizing 1D $IrO_2$ stripes using *in-plane* oriented superlattices of $Sr_2IrO_4$ and the wide bandgap insulator $LaSrGaO_4$ ($E_g$ = 3.8 eV) (Figure 1b and Figure S1 in supporting information). The 2D layered iridate $Sr_2IrO_4$ exhibits a unique spin-orbit coupled $J_{eff}$ = 1/2 Mott insulating state, which is not present in its three-dimensional metallic counterpart $SrIrO_3$.[16-19] Recent studies on 2D $Sr_2IrO_4$ also have revealed interesting phenomena such as the presence of excitonic quasiparticles and indications of new superconducting ground states for doped samples.[20-23] Although expected to form intriguing electronic and magnetic ground states in one dimension, there are no previously studied or discovered 1D iridate systems to date. To synthesize the 1D superlattice structures, we have used a customized pulsed laser deposition (PLD) system in order to optimize monolayer-controlled deposition of $Sr_2IrO_4$ and $LaSrGaO_4$ (see supporting information and Ref. [24] for detailed growth methods and conditions). Using structural and electronic characterizations of x-ray diffraction, high-resolution scanning transmission electron microscopy (STEM), optical spectroscopy, and resonant inelastic x-ray scattering, we have confirmed that both 1D structural and electronic confinement is achieved for this system. Our experimental observations are also consistent with the calculations of density functional theory. These results imply that this method can be extended to any 2D layered material and thereby allow for tunability to 1D quantum structures previously disregarded in experimental one-dimensional science.

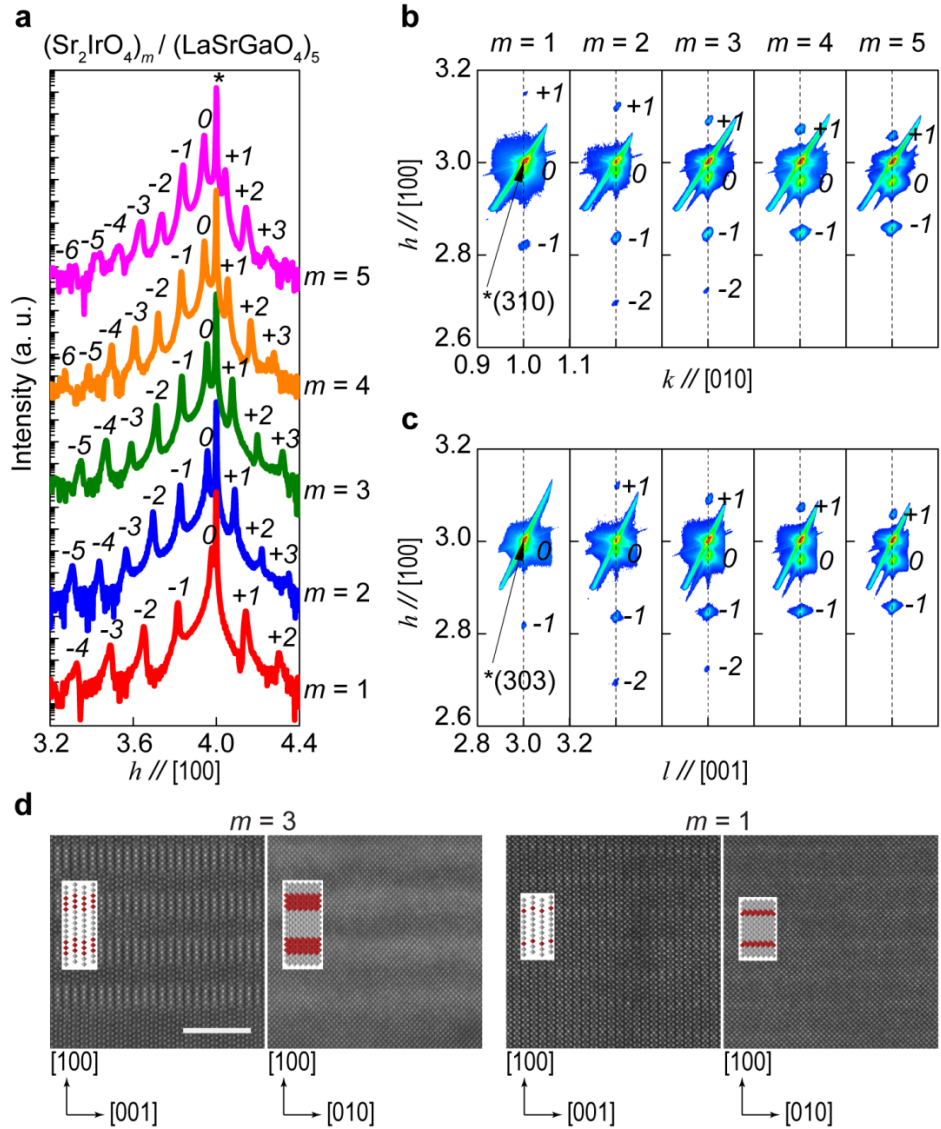

**Figure 2.** X-ray diffraction scans and Z-contrast STEM data of $(Sr_2IrO_4)_m/(LaSrGaO_4)_5$ superlattices. a) $2\theta$-$\omega$ x-ray diffraction scans of $(Sr_2IrO_4)_m/(LaSrGaO_4)_5$ superlattices (labeled accordingly to the right). The $LaSrGaO_4$ (400)-diffraction peaks are indicated by an asterisk (*). The central superlattice peak is the zeroth order Bragg diffraction peak (indexed as *0*), and the superlattice satellite peaks are indexed relative to this peak. Two x-ray reciprocal space maps are taken for each superlattice sample about b) the (310)- and c) the (303)-reflections of $LaSrGaO_4$ (*) to obtain complete in-plane strain information. d) Z-contrast STEM images of *m* = 3 (left pair) and *m* = 1 (right pair) superlattices. Two cross-sections are shown for each sample: the *ac*-plane (left scan of each pair) and the *ab*-plane (right scan of each pair). The brightest dots are Ir ions and the inset schematics employ the same color scheme as in Figure 1b. Note that the $IrO_2$ stripes run along the *b*-axis ([010]-direction), that is the out of page direction in the left image

and the horizontal direction in the right image of each sample. All images have the same scale bar (leftmost image) of 5 nm.

The 1D stripe-structures of our superlattice samples have been confirmed by x-ray diffraction and high-resolution STEM. Figure 2a shows clear periodic superlattice peaks in the $2\theta$-$\omega$ x-ray diffraction scans. Since each superlattice unit cell consists of $m$ monolayers of $Sr_2IrO_4$ and 5 monolayers of $LaSrGaO_4$, i.e. $(Sr_2IrO_4)_m/(LaSrGaO_4)_5$ ($m = 1 - 5$), the superlattice diffraction peak periodicity $\Delta h$ shows excellent agreement to $1/(m + 5)$. For strain information, x-ray reciprocal space maps have been taken near the (310)- and (303)- reflections of $LaSrGaO_4$ for the $ab$- (Figure 2b) and $ac$-planes (Figure 2c), respectively. The vertical alignment of the superlattice peaks with the substrate reflections indicates all superlattices are coherently strained along both directions, i.e. $b$- and $c$-axes. STEM further confirms that the 1D superlattice structures are successfully formed (Figure 2d). The expected half-unit cell offset (zig-zag shape) and Sr-O rock salt separation between neighboring $IrO_2$ stripes along the $c$-axis is readily apparent in the $ac$-plane scans of the $m = 3$ and $m = 1$ samples. Along the $b$-axis, the $IrO_2$ stripes are coherent; however, the stripe boundaries in this planar view are rather indeterminate due to the inter-stripe half unit cell offset between each planar layer. These observations provide convincing evidence for the successful creation of 1D $IrO_2$ stripe-structures from the 2D $Sr_2IrO_4$.

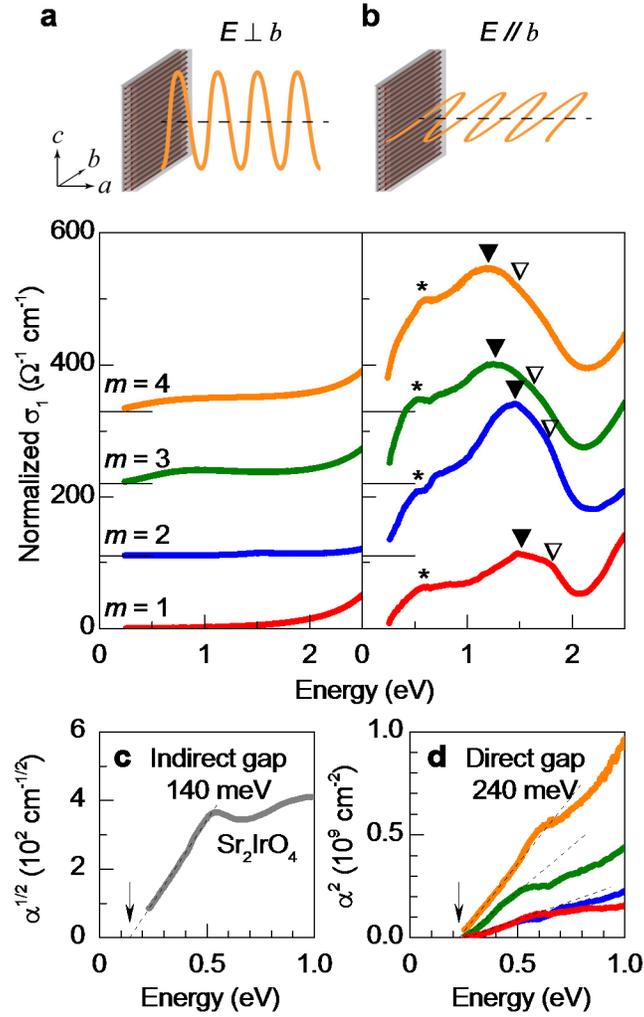

**Figure 3.** Optical spectra of $(Sr_2IrO_4)_m/(LaSrGaO_4)_5$ superlattices. Above each plot, an experimental depiction of the two sets of incident photons with linear polarization ($E$) a) perpendicular and b) parallel to the $IrO_2$ stripe-direction ($b$-axis) for $m = 1$ (red), 2 (blue), 3 (green), and 4 (orange) superlattices. All optical conductivity spectra $\sigma_1$ are normalized by the number of Ir monolayers per superlattice unit cell for comparison, and $m = 2$, 3, and 4 spectra are vertically shifted for clarity. The strong optical anisotropy, i.e. the absence of the low energy absorption peaks when $E$ is perpendicular to the 1D $IrO_2$ stripes, confirms the low-dimensional electronic confinement. In the $m = 4$ superlattice spectra, the two peaks indicated by * and ▼ closely match the peak positions of 2D $Sr_2IrO_4$ crystals[15]. As the dimensionality decreases, several notable features include: the change in electronic band and optical gap nature from c) an indirect gap (band edge linear with $\alpha^{1/2}$) to d) a direct gap (band edge linear with $\alpha^2$), the appearance of van Hove singularities, the distinct emergence of a higher energy peak marked by ▽, and a blue-shift of the ▼ peak.

Linearly polarized optical spectroscopy shows clear anisotropic characteristics and 1D electronic confinement of the stripe-structures.  The directional-dependent absorption spectra are obtained by linearly polarizing the incident photons such that the electric field $E$ is perpendicular (parallel) to the 1D $IrO_2$ stripe direction along the $b$-axis, as shown by the graphic in Figure 3a (Figure 3b).  The optical conductivity spectra ($\sigma_1(\omega)$) are obtained by taking a Kramers-Kronig transformation of the absorption spectra $\alpha(\omega)$ (Figure S4 in the Supporting Information) and are normalized by the number of Ir monolayers per superlattice unit cell, i.e. $m/(m + 5)$ for the $(Sr_2IrO_4)_m/(LaSrGaO_4)_5$ superlattices.  When the polarization is parallel to the 1D $IrO_2$ stripe direction, i.e. $E // b$ (Figure 3b), there are broad peaks in the optical spectra.  The finite peaks appear due to the Ir-Ir inter-site optical transitions in the $J_{eff} = 1/2$ band.[21, 25, 26]  Note that the peaks are significantly reduced (Figure 3a) when the polarization is perpendicular to the $IrO_2$ stripe direction, i.e. $E \perp b$.  If the $IrO_2$ stripes are truly 1D, this can be understood intuitively since no Ir-Ir inter-site optical transitions are possible perpendicular to the $b$-axis.  Hence, the observation of optical anisotropy confirms that the 1D $IrO_2$ stripes are also confined *electronically*.  As $m$ decreases, i.e. the samples approach one dimension, the spectral weight in the absorption is reduced (Figure S4 in the Supporting Information) and the indirect gap observed in 2D $Sr_2IrO_4$ (Figure 3c) becomes a direct gap for the 1D superlattices (Figure 3d).  New sharp features also appear in the spectra (marked as $\nabla$ in Figure 3b).  The sharp features present in the optical spectra are attributed to van Hove singularities, the appearance of which is contingent upon the low dimensionality of the crystal.[27]  The low dimensionality also contributes to an enhanced effective electron-correlation $U_{eff}$, which is responsible for the blue-shift of the optical transitions (Figure S4 in the Supporting Information).

In order to reveal the elementary excitation dynamics of spin/orbital degrees of freedom exhibited by the 1D $IrO_2$ stripe-structures, we have taken resonant inelastic x-ray scattering (RIXS) spectra (For measurement geometry, see Figure S6 in the Supporting Information). In the case of 2D $Sr_2IrO_4$, RIXS spectra reveal a magnon produced by antiferromagnetic Heisenberg spin below 0.2 eV as well as dispersive excitonic quasiparticles of $J_{eff}$ orbitals (spin-orbit excitons) above 0.4 eV (Figure 4a and Ref. [20]). For the 1D superlattice, the $m = 1$ RIXS spectra reveal many unique features not present in its 2D $Sr_2IrO_4$ counterpart (Figure 4b and Figure 4c). The spin excitation energy of the 1D $IrO_2$ stripes (~0.2 eV) is higher than that of 2D $Sr_2IrO_4$ (~0.06 eV), which implies that the 1D $IrO_2$ stripes have higher exchange interactions due to the localized character of the 1D spin structure. In the region of orbital excitations, a resolution-limited spin-orbit exciton at 0.6 eV is discovered. This is interpreted as arising from the same $J_{eff}$ orbital transitions of the spin-orbit exciton in bulk $Sr_2IrO_4$— with a notable exception of broad dispersive excitations within the vicinity of the 1D spin-orbit exciton. These broad excitations disperse towards higher energy as the momentum transfer approaches $\pi$ (Figure 4b). Note that broad excitations in bulk $Sr_2IrO_4$ originate from the electron-hole continuum and damped orbital excitons. Thus, the broad dispersive excitations in the 1D $IrO_2$ stripe RIXS spectra are consistent with the optical spectroscopy data (Figure 3b), which reveal the existence of a largely reduced electron-hole continuum. Hence, the observed broad orbital excitations reflect the deconfined character of orbitals in the 1D structure.

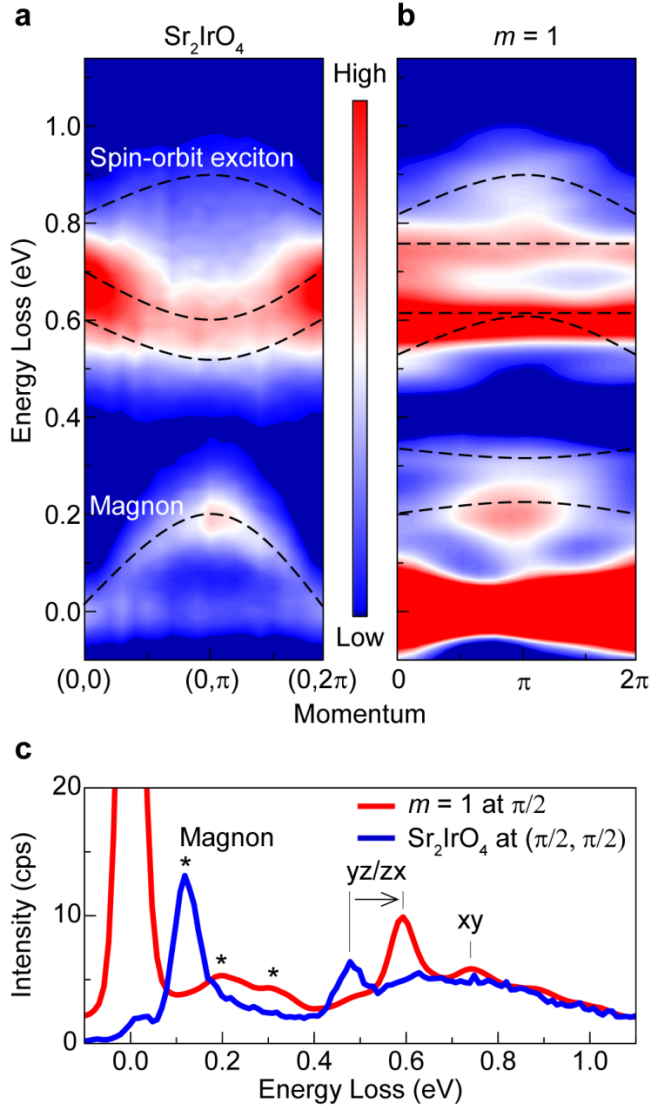

**Figure 4.** Intensity contour plots of RIXS spectra for a) 2D $Sr_2IrO_4$ crystal along the (0, 0) to (0, $2\pi$) direction of its 2D Brillouin zone and b) 1D $(Sr_2IrO_4)_1/(LaSrGaO_4)_5$ superlattice along the 0 to $2\pi$ direction (i.e. parallel to the *b*-axis) of its 1D Brillouin zone. Note that the low-energy magnon branch and the dispersion of spin-orbit excitons are shown below 0.2 eV and above 0.4 eV, respectively, and dashed lines act as visual guides for qualitatively contrasting the dimensionality effect on these dispersive features. c) RIXS spectra of 1D $(Sr_2IrO_4)_1/(LaSrGaO_4)_5$ superlattice (red) and 2D $Sr_2IrO_4$ crystal (blue) at comparable momentum transfers, i. e. $\pi/2$ and $(\pi/2, \pi/2)$, respectively. Distinct excitations of the 1D superlattice include the following: broad spin-excitations of the 1D superlattice below 0.4 eV are in sharp contrast from the resolution-limited magnon peak of 2D $Sr_2IrO_4$ (indiated by * in c); the 1D superlattice shows a quasiparticle spin-orbit exciton at ca. 0.6 eV, of which energy is 100 meV higher than that of 2D $Sr_2IrO_4$; and broad orbital excitations also show different dispersive behaviors from 2D counterpart. The strong intensity around zero energy loss in the 1D superlattice spectra is mostly from elastic scattering by $LaSrGaO_4$ layers.

To supplement the understanding of the experimental data, we have performed density functional theory (DFT) calculations for the 1D $(Sr_2IrO_4)_1/(LaSrGaO_4)_5$ stripe-structure.  As compared to its 2D $Sr_2IrO_4$ counterpart (Figure S5a in the Supporting Information), the spin-orbit split $J_{eff} = 1/2$ band appears quite flattened in the 1D energy dispersion (Figure S5b in the Supporting Information), indicating the presence of a very localized $J_{eff} = 1/2$ state.  As a consequence, the indirect gap between the $J_{eff} = 1/2$ state in 2D $Sr_2IrO_4$ becomes a direct gap in its 1D counterpart, which is in excellent agreement with the indirect-to-direct gap phase transition observed in the optical spectra (Figures 3c and 3d, respectively).  The localization of the $J_{eff} = 1/2$ state should also induce additional sharp optical transitions since optical spectroscopy measures the joint density of states between occupied and unoccupied bands.  This result is consistent with the emergence of the van Hove singularity, i.e. sharp peak $\nabla$, in the optical conductivity spectra of the superlattices (Figure 3b).  The excellent agreement of these 1D characters with the experimentally obtained optical conductivity spectra verifies the successful electronic confinement of our $(Sr_2IrO_4)_m/(LaSrGaO_4)_5$ stripe-structures.

This approach of creating 1D quantum-stripe systems provides an avenue for exploring emergent phenomena of low dimensional physics.  For instance, superconducting 1D nano-stripes can be created by using an $a$-axis oriented superlattice of the layered cuprates such as $(La,Sr)_2CuO_4$.  Although metallic ground states in true 1D materials are hard to stabilize due to structural instabilities such as the Peierls transition, the striped structures can be used to discover new phase transitions by tuning the system's dimensionality from two to one dimension(s).[28] Another intriguing feature of low dimensional systems is fractionalization of spin and charge degrees of freedom (e.g. spinons and excitons).  Such excitonic effects are expected to be enhanced due to their spatial decoupling in 1D systems.

In summary, we have shown a generalizable superlattice approach of a continuous dimensional control between two-dimensional (2D) layered oxides and 1D quantum stripes. We demonstrated its successful application on the highly correlated, 2D $Sr_2IrO_4$. In our superlattice structures, we have observed the optical transitions of electrons between neighboring Ir atoms to be confined to the 1D $IrO_2$ stripe direction, which emulates ideal 1D behavior. Spin and orbital excitations observed in resonant inelastic x-ray scattering also suggest enhanced spin exchange interactions and confined orbital excitations in the 1D $IrO_2$ stripes as compared to 2D $Sr_2IrO_4$. This 1D superlattice method can be readily adopted for unveiling 1D phenomena in a variety of two-dimensional materials.

**Acknowledgements**

We acknowledge the support of National Science Foundation (NSF) grant DMR-1454200 for sample synthesis and characterizations. J.W.B acknowledges the support of NSF grant DMR-1262261 for infrared spectroscopy. Research at the University of Toronto was supported by the NSERC of Canada and the Center for Quantum Materials at the University of Toronto. Computations were mainly performed on the GPC supercomputer at the SciNet HPC Consortium. SciNet is funded by the Canada Foundation for Innovation under the auspices of Compute Canada; the Government of Ontario; Ontario Research Fund—Research Excellence; and the University of Toronto. This research used resources of the Advanced Photon Source, a U.S. Department of Energy (DOE) Office of Science User Facility operated for the DOE Office of Science by Argonne National Laboratory under Contract No. DE-AC02-06CH11357.

## Supporting Information

**Engineering One-Dimensional Quantum Stripes from Superlattices of Two-Dimensional Layered Materials**

*John H. Gruenewald, Jungho Kim, Heung Sik Kim, Jared M. Johnson, Jinwoo Hwang, Maryam Souri, Jasminka Terzic, Seo Hyoung Chang, Ayman Said, Joseph W. Brill, Gang Cao, Hae-Young Kee, and Sung S. Ambrose Seo\**

## Methods

## Experimental Section

*Sample fabrication and characterization*: The $(Sr_2IrO_4)_m/(LaSrGaO_4)_5$ ($m = 1 - 5$) superlattices are grown using a custom-built pulsed laser deposition system with in situ reflection high-energy electron diffraction (RHEED) and in situ optical spectroscopic ellipsometry. We have used $a$-axis oriented $LaSrGaO_4$ (100) as both the substrate and insulating sublayers. The samples in this study are 30 superlattice-unit-cells thick, and each superlattice unit cell consists of $m$ monolayers of $Sr_2IrO_4$ and 5 monolayers of $LaSrGaO_4$. In order to achieve a controlled monolayer growth, it is important to maintain a slow growth rate of $Sr_2IrO_4$ and $LaSrGaO_4$ layers, ca. 200 laser pulses per monolayer. The growth parameters of the deposition require an oxygen partial pressure ($P_{O2}$) of 10 mTorr, a substrate temperature of 700 °C, and laser (KrF excimer, $\lambda = 248$ nm) fluence of 1.2 J/cm$^2$. The superlattice growth is monitored using RHEED (see Figure S2 in the Supporting Information), and the deposited monolayers of both $LaSrGaO_4$ and $Sr_2IrO_4$ show distinct axis-dependent diffraction patterns during the entire deposition, which indicate the proper structural confinement required for creating the 1D $IrO_2$ stripe-structures.

The x-ray diffraction $2\theta$-$\omega$ scans and RIXS were measured using the Advanced Photon Source at Argonne National Laboratory. The x-ray reciprocal space maps were taken using a Bruker D8 Advance. The polarization dependent absorption spectra were taken using a Fourier-transform infrared spectrometer in the photon energy region of 50 meV–0.6 eV and a grating-type spectrophotometer in the range of 0.5–6 eV.

*DFT Calculations:* From the DFT calculations (Figure S5 in the Supporting Information), we find the tight binding parameters between the nearest-neighboring $t_{2g}$ orbitals as $H_{t_{2g}} =$

$$\begin{pmatrix} t_3 & & \\ & t_1 & t_2 \\ & -t_2 & t_1' \end{pmatrix},$$ where $t_1$ = -0.330 eV, $t_1'$ = -0.294 eV, and $t_3$ = -0.082 eV, which are hopping integrals between the $d_{xy}$, $d_{yz}$, and $d_{xz}$ orbitals respectively (with the convention x//$a$, y//$b$, and z//$c$ in Figure 1b). Due to the small but existing octahedral rotations along the $b$-direction, $t_2$ = 0.010 eV is also introduced. The largest interchain hopping term $t_p$ is ca. -0.02 eV, and the dispersion along the $\boldsymbol{a}$-direction is vanishingly small. From this analysis, the system is quasi-1D at room temperature where the optical spectra are collected. Due to the strong spin-orbit coupling (SOC), the band obtained at zero temperature consists mostly of $J_{eff}$ = 1/2 as shown in Figure S5 in the Supporting Information. Antiferromagnetic ordering with the spins parallel to the $c$-axis is also found. In the large $U$ limit, we find that the effective spin model maps to an XXZ model with small anisotropy and tiny interchain Heisenberg term $J_p$, with $J_x \cong J_z$ = 0.20 eV, $J_y$ = 0.22 eV, and $J_p$ = 5 meV assuming $U$ = 2 eV and the Hund's coupling $J_H$ = 0.4 eV (Ref. [1]). Additionally, due to the small octahedral rotations, the Dzyaloshinskii-Moriya (DM) vector with |DM| $\cong$ 10 meV parallel to the $\boldsymbol{b}$-direction is also introduced.[2, 3] Note that the presence of bond-direction anisotropy of $J_y$-$J_{x,z}$ = 20 meV is larger than the magnitude of the DM vector and thus favors spin alignment parallel to the chain direction.

The DFT results in this study were obtained by using two different density functional theory codes: OPENMX and Vienna ab-initio Simulation Package (VASP).[4-6] For the structural optimization, we used the projector-augmented wave potentials and the PBEsol generalized gradient approximation as implemented in VASP.[7]  A 4×2×1 Monkhorst-Pack grid and 400 eV of plane-wave energy cutoff were used for the momentum space sampling and the basis set, respectively.  The force criterion was $10^{-3}$ eV/Å.  For the electronic structure calculations with SOC and on-site Coulomb interaction, OPENMX code, which is based on the linear-combination-of-pseudo- atomic-orbital basis formalism, was used.[4]  A noncollinear DFT scheme and a fully relativistic J-dependent pseudopotential were used to treat SOC, and the parametrization of Perdew and Zunger for the local density approximation (LDA) was chosen for the exchange-correlation functional.[8, 9]  An energy cutoff of 300 Ry was used for the real-space integration.  On-site Coulomb interaction was treated via a simplified DFT+U formalism implemented in OPENMX code, and $U_{\text{eff}} \equiv U - J = 2.0$ eV was employed for Ir $d$-orbital in our DFT+SOC+U calculations.[10, 11]  Using maximally localized Wannier orbital formalism implemented in OPENMX code, we obtained the Ir $t_{2g}$ hopping parameters for estimating the 1D $J_{\text{eff}} = 1/2$ spin Hamiltonian.[12-14]

**Figure S1. Schematic detailing the atomic and geometric configurations of Sr₂IrO₄ and LaSrGaO₄ unit cells.** The 2D planes in $Sr_2IrO_4$ and $LaSrGaO_4$ consist of an array of octahedra, which each corner-share oxygen atoms in the *ab*-plane and centrally contain the transition metal Ir or Ga, respectively. The 2D planes are each separated by a rock salt layer of SrO for $Sr_2IrO_4$ and (La,Sr)O for $LaSrGaO_4$, and adjacent planes are each offset by a half unit cell. For our 1D *m* = 1 structure, one monolayer (a half unit cell) of $Sr_2IrO_4$ is deposited along the *a*-axis followed by 5 monolayers (four and a half unit cells) of $LaSrGaO_4$.

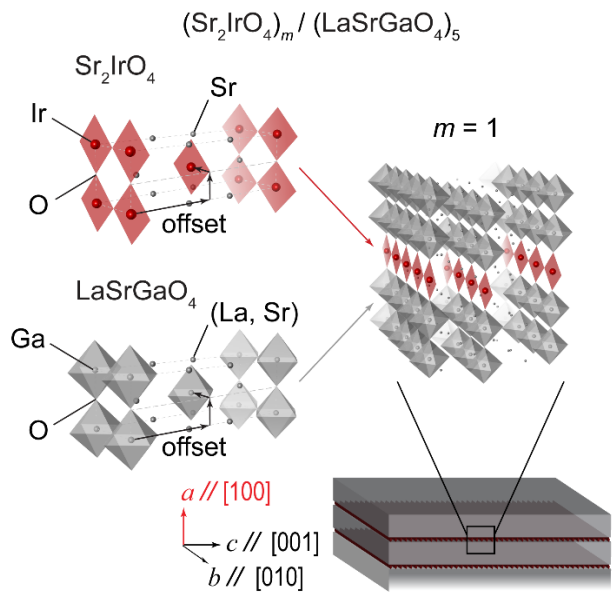

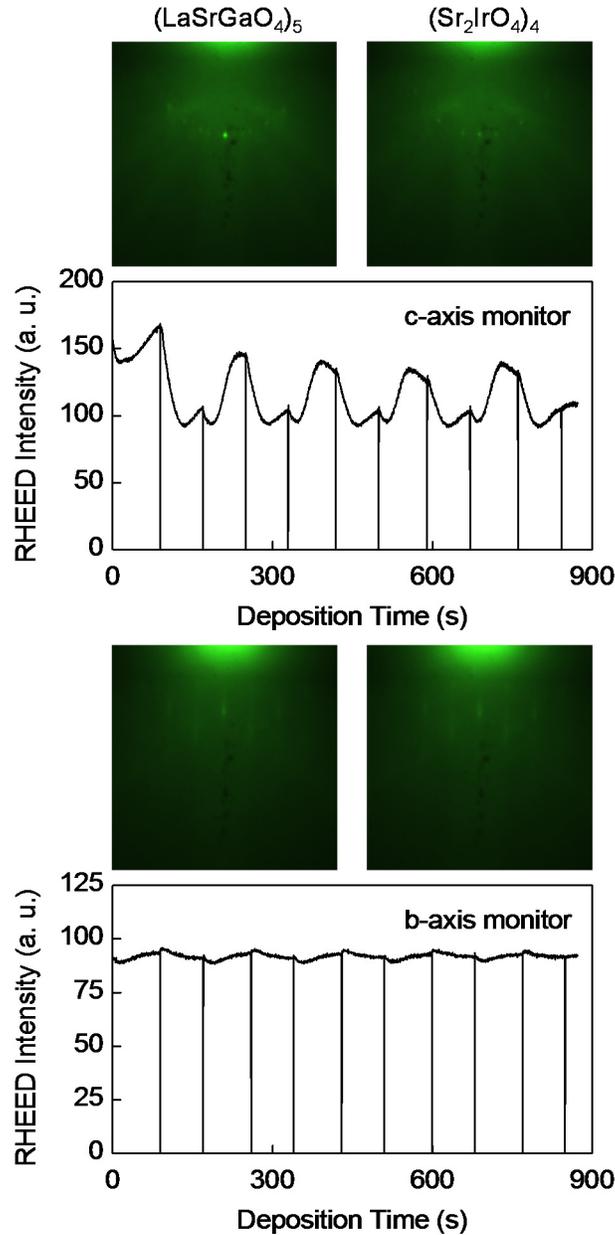

**Figure S2.** **Reflection high-energy electron diffraction (RHEED) monitor during _m_ = 4 growth.** The in-plane anisotropy of the $(Sr_2IrO_4)_4/(LaSrGaO_4)_5$ superlattice induces distinct axis dependent RHEED diffraction patterns and intensity fluctuations for monitoring the _c_-axis (top panel) and _b_-axis (bottom panel). Before growing the superlattice structure, a few monolayers of $LaSrGaO_4$ are deposited to act as a buffer layer. The RHEED intensity and diffraction patterns remain relatively stable during the growth, indicating the superlattice retains structural integrity during the deposition.

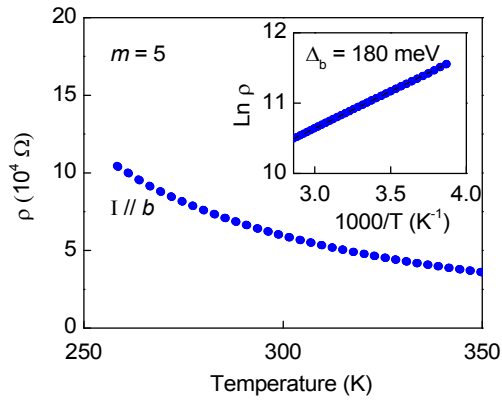

**Figure S3.** **Arrhenius plot of the resistivity for (Sr$_2$IrO$_4$)$_5$/(LaSrGaO$_4$)$_5$.** Resistivity is taken of $m = 5$ from 300 K to just above 250 K, and the charge gap $\Delta_b$ is extracted from a fit to the Arrhenius equation: $\rho\,(T) = \rho_0 \exp(\Delta/2k_B T)$, as shown in the inset. The blue curve in the plot is the resistivity parallel to the IrO$_2$ stripe direction along the $b$-axis. The other superlattice structures ($m = 1 - 4$) are too insulating for reliable resistivity measurements.

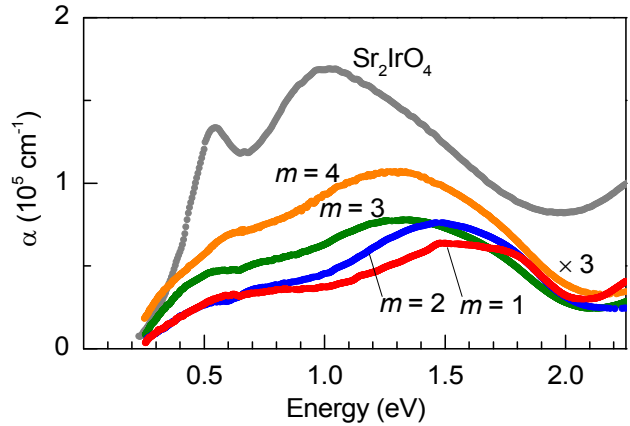

**Figure S4.** **Linearly polarized optical absorption spectra of (Sr₂IrO₄)ₘ/(LaSrGaO₄)₅ superlattices (*m* = 1 – 4).** The absorption spectra are taken at normal incidence and polarized such that the AC electric field $E$ is parallel to the $IrO_2$ stripe direction (*b*-axis), as schematically depicted in Figure 3b. In comparing with $Sr_2IrO_4$ (grey), the superlattice absorption monotonically decreases as 1D is approached from *m* = 4 to *m* = 1, and the blue shift of the central absorption peak is also readily apparent. For visual clarity, the absorption spectra for *m* = *1, 2, 3, 4* have been multiplied by a factor of three.

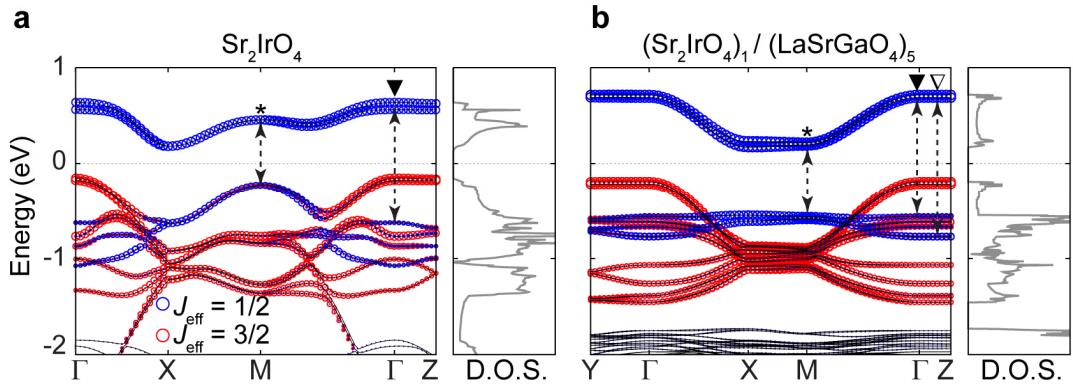

**Figure S5.** **Band structures from DFT calculations and the density of states (DOS).** a) The calculated electronic band structure (left) and DOS (right) of 2D $Sr_2IrO_4$. The $J_{eff} = 1/2$ and $J_{eff} = 3/2$ states are distinguished between blue and red circles, respectively. The optically allowed inter-band $J_{eff} = 1/2$ transitions [20] are indicated by dashed lines, and the labels * and ▼ correspond to the observed peaks in the optical conductivity spectra in Figure 3b. b) The electronic band structure and DOS of 1D $(Sr_2IrO_4)_1/(LaSrGaO_4)_5$ superlattice show the flat dispersion due to dimensional confinement. An indirect-to-direct gap transition is consistent with experimental data (Figure 3c and Figure 3d). Sharp features (van Hove singularities) in the DOS and an additional optically allowed $J_{eff} = 1/2$ transition (▽) are in agreement with the experimental spectra in Figure 3b.

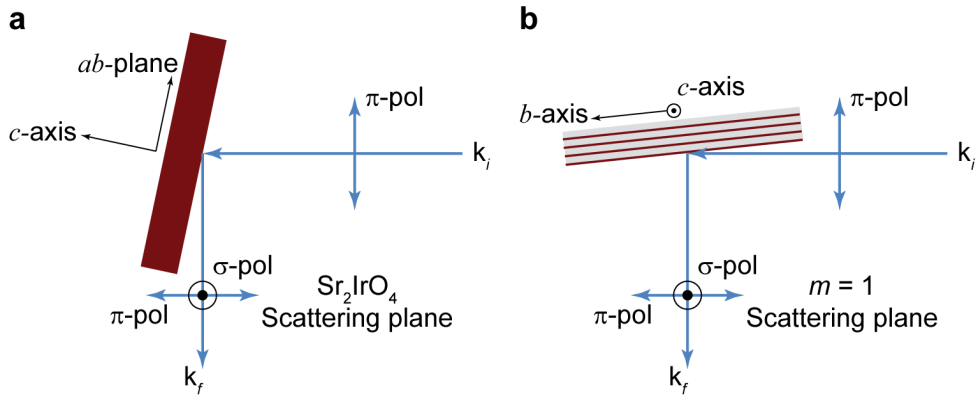

**Figure S6.** **Measurement geometry of resonant inelastic scattering measurements (RIXS) with respect to Sr$_2$IrO$_4$ and the 1D $m$ = 1 superlattice.** The incident-photon is polarized along a) the ab-plane for bulk Sr$_2$IrO$_4$ and along b) the a-axis for the 1D $m$ = 1 superlattice. The scattered-photon polarization has components both parallel to the incident polarization ($\pi$-polarization) and perpendicular to the scattering plane ($\sigma$-polarization). In this geometry, the in-plane spin components (*ab*-plane of bulk Sr$_2$IrO$_4$ and *b*-axis of the 1D $m$ = 1 superlattice) and the out-of-plane spin components (*ab*-plane of bulk Sr$_2$IrO$_4$ and *c*-axis of 1D $m$ = 1 superlattice) are probed. The scattered-photon polarization is used to probe the orbital lobes of the *c*-axis and in-plane directions for both samples. The geometry in a) is used to maximize the RIXS matrix elements for the transition between |J$_{eff}$ = 1/2, J$_z$ = 1/2> (= |xy,+> + |yz,-> + i|zx,->) and |J$_{eff}$ = 3/2, J$_z$ = ±3/2> (=|yz,±> ± i|zx,±>) [19]. The resolution-limited orbital exciton peak at 600 meV of the 1D $m$ = 1 superlattice can be assigned as from the same orbital transition. The same energy resolution of 30 meV is used for both measurements, and incident- and scattered-photon polarizations are also similar in both cases.